**scientific** reports

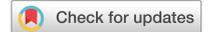

OPEN

# A universal model for the Lorenz curve with novel applications for datasets containing zeros and/ or exhibiting extreme inequality

Thitithep Sitthiyot[1✉] & Kanyarat Holasut[2]

Given that the existing parametric functional forms for the Lorenz curve do not fit all possible size distributions, a universal parametric functional form is introduced. By using the empirical data from different scientific disciplines and also the hypothetical data, this study shows that, the proposed model fits not only the data whose actual Lorenz plots have a typical convex segment but also the data whose actual Lorenz plots have both horizontal and convex segments practically well. It also perfectly fits the data whose observation is larger in size while the rest of observations are smaller and equal in size as characterized by two positive-slope linear segments. In addition, the proposed model has a closed-form expression for the Gini index, making it computationally convenient to calculate. Considering that the Lorenz curve and the Gini index are widely used in various disciplines of sciences, the proposed model and the closed-form expression for the Gini index could be used as alternative tools to analyze size distributions of non-negative quantities and examine their inequalities or unevennesses.

The distributions of sizes vary significantly in both nature and society. Extreme inequalities in size distributions are also not unusual. In nature, based on the datasets used in Newman's study[1], the share of the top 10% of earthquake intensity is equal to 16% of total share of earthquake intensity while the share of the top 10% of solar flare intensity accounts for 85% of total share of solar flare intensity. In addition, the top 10% of mammal species' body mass has a share of 99% of total share of mammal species' body mass[2]. The degree of metabolic network of the bacterium *Escherichia coli* exhibits a similar pattern in that the share of the top 10% of metabolic network accounts for 99.9% of total share of metabolic network[3]. In society, according to the data from the American Federation of Labor and Congress of Industrial Organizations (AFL-CIO)[4], the top 10% of compensation of chief executive officers (CEOs) has a share of 27% of total share of CEOs' compensation whereas the data on salary of professional athletes[5] show that the top 10% of professional women tennis players' salary has a share of 76% of total salary share of professional women tennis players. Furthermore, based on the inter-state war data[6], the share of the top 10% of war intensity as measured by the number of deaths per battle accounts for 91% of total share of the number of deaths per battle.

To analyze the distributions of sizes and examine the inequalities in size distributions, a tool that has been commonly used for more than a century is the Lorenz curve. It was originally developed by an American economist named Max O. Lorenz[7] as a method for measuring wealth concentration. The Lorenz curve depicts a graphical relationship between the cumulative normalized rank of population from the poorest to the richest (the abscissa) and the cumulative normalized wealth held by these population from the poorest to the richest (the ordinate). The application of the Lorenz curve is not limited to economics, however. According to Eliazar and Sokolov[8], the use of the Lorenz curve has grown beyond economics and reached various disciplines of sciences.

There are three popular methods that could be used to estimate the Lorenz curve. They are: (1) interpolation techniques (2) specifying a statistical distribution of size and deriving the corresponding Lorenz curve and (3) specifying a parametric functional form for the Lorenz curve. Given that the interpolation techniques underestimate inequality unless the individual data on size are available and no single statistical distribution has proved to be adequate for representing the entire size distribution[9], numerous studies have proposed a variety of parametric functional forms in order to directly approximate the Lorenz curve[9–34].

[1]Department of Banking and Finance, Faculty of Commerce and Accountancy, Chulalongkorn University, Mahitaladhibesra Bld., 10th Fl., Phayathai Rd., Pathumwan, Bangkok 10330, Thailand. [2]Department of Chemical Engineering, Faculty of Engineering, Khon Kaen University, Mittapap Rd., Muang District, Khon Kaen 40002, Thailand. ✉email: thitithep@cbs.chula.ac.th

According to Dagum[35], a good parametric functional form for estimating the Lorenz curve has to be able to describe the distributions of sizes via the changes in parameter values. The specified functional form should also provide a good fit for the entire range of size distribution since all observations are relevant for an accurate measurement of inequality or unevenness. Sitthiyot and Holasut[34] note that while many popular parametric functional forms for the Lorenz curve do not have a closed-form expression for the Gini index, making it computationally inconvenient to calculate since they require the valuation(s) of the beta function[11–13,19], or the beta and the gamma functions[24], or the confluent hyper-geometric function[17], a good functional form for the Lorenz curve therefore should have an explicit mathematical solution for the Gini index[35]. In addition, Dagum[35] suggests that a good parametric functional form should employ the smallest possible number of parameters for adequate and meaningful representation with well-defined meanings. While three- or four- parameter functional form implies a loss in simplicity, a functional form that fits the empirical data well with an associated inequality measure such as the Gini index usually requires more than two parameters. Furthermore, Dagum[35] notes that, from a viewpoint of computational cost and the acceptance of the specified functional form in applied sciences, a simple method of parameter estimation is always an advantage.

Thus, finding a parametric functional form that satisfies the aforementioned properties of a good parametric functional form for the Lorenz curve as suggested by Dagum[35] is a theoretical and practical challenge. In addition, given the fact that sizes of events or things that occur in nature and society could contain zeros such as the magnitude of earthquake intensity based on the Richter scale, the number of connections in metabolic network in living organisms, and the degree of war intensity as measured by the number of deaths per battle, to our knowledge, no study has proposed a parametric functional form that takes this possibility into account by allowing the Lorenz curve to have a horizontal-line segment in addition to a typical convex segment. Empirically, it is also possible that one observation has a larger size compared to the rest of observations whose sizes are smaller and more or less equal. For example, in nature, queen termites could live for 20 years whereas worker termites live only a few weeks to months[36]. In society, one person could have a majority of income share while the income shares of the rest of population are more or less equal. Based on these two examples, a good parametric functional form for the Lorenz curve that fits extreme inequality in the distributions of age of termites and income of population should have two positive-slope linear segments. Also, to our knowledge, there is no existing parametric functional form for the Lorenz curve whose performance is up to this task.

To address the key issues with regard to the existing parametric functional forms for estimating the Lorenz curve as discussed above and to fill the gap in the literature on the Lorenz curve, this study introduces a universal parametric functional form for the Lorenz curve (proposed model) that has a closed-form expression for the Gini index. Our proposed model has four parameters. It comprises a linear function and a linear combination of two convex functions which are the exponential function and the functional form implied by Pareto distribution. According to Ogwang and Rao[21], a linear combination is a way to circumvent an important drawback of traditional parametric functional forms for the Lorenz curve which is the lack of satisfactory fit over the entire range of a given size distribution. Note that the mixture of the exponential function and the functional form implied by Pareto distribution accounts for the convex segment of the Lorenz curve. The linear function is incorporated in order to characterize the horizontal-line segment of the Lorenz curve where a certain number of observations have a size of zero and/or to represent the positive-slope linear segments of the Lorenz curve where the size of one observation is larger than the sizes of the rest of observations which are smaller and more or less equal.

To demonstrate the performance of our proposed model for the Lorenz curve, the empirical data on sizes of events or things, occurring in nature and society, from across scientific disciplines are used. They are earthquake intensity, solar flare intensity, mammal species' body mass, metabolic network of the bacterium *Escherichia coli*, CEOs' compensation, salary of professional women tennis players, and inter-state war intensity. These datasets are publicly available and almost all of them can be accessed from the sources mentioned earlier. Note that the data on the earthquake intensity, the metabolic network of the bacterium *Escherichia coli*, and the intensity of inter-state war contain numerous observations whose sizes are equal to zero. In addition, a hypothetical dataset is created in order to illustrate how the proposed model could be used to fit the distribution of size where one observation has a larger size compared to the others which have smaller and equal size. In addition, this study compares the performance of the proposed model to that of Sarabia et al.[24] (SCS model) which, according to Tanak et al.[37], is considered the best performer among a number of different functional forms for the Lorenz curve in fitting to the data. Although the SCS model has been shown to outperform a number of different parametric functional forms for the Lorenz curve, it has three parameters. Thus, to level the playing field, a parametric functional form that contains four parameters developed by Sarabia[23] (S model) is also employed for the performance comparison. The main reason that we choose the S model because it has been demonstrated to fit the data better than other well-known parametric functional forms for the Lorenz curve such as Chotikapanich[9], Kakwani and Podder[10], Rasche et al.[13], and Arnold[16]. Given that the Lorenz curve and the Gini index are extensively used in numerous disciplines of sciences, our proposed model for the Lorenz curve with a closed-form expression for the Gini index could be used as an alternative tool for analyzing size distributions of non-negative quantities and examining inequalities or unevenness.

## Methods

Let $x$ be the cumulative normalized rank of size from the smallest to the largest and $y$ be the cumulative normalized size from the smallest to the largest, where $0 \leq x \leq 1$ and $0 \leq y \leq 1$. In addition, let $\delta, \rho, \omega,$ and $P$ denote parameters, where $0 \leq \delta < 1, 0 \leq \rho \leq 1, 0 \leq \omega \leq 1,$ and $P \geq 1$. While there is a vast family of existing and already known parametric functional forms for estimating the Lorenz curve that could be used in combination by assigning a weight between 0 and 1 to each functional form such that the sum of all weights is equal to 1[34], our proposed model is characterized by three functions which are as follows:

Linear function:

$$y(x) = \left( \frac{2}{P+1} \right) * \left( \frac{x-\delta}{1-\delta} \right), \tag{1}$$

given

(1)  $y(x) = 0$ when $x - \delta < 0$,
(2)  $y(x) = \left( \frac{2}{P+1} \right) * \left( \frac{x-\delta}{1-\delta} \right)$ when $0 \le x - \delta < 1$,
(3)  $y(x) = 1$ when $x = 1$.

Exponential function:

$$y(x) = \left( \frac{x-\delta}{1-\delta} \right)^P, \tag{2}$$

given

(1)  $y(x) = 0$ when $x - \delta < 0$,
(2)  $y(x) = \left( \frac{x-\delta}{1-\delta} \right)^P$ when $0 \le x - \delta \le 1$.

Functional form implied by Pareto distribution:

$$y(x) = 1 - \left( 1 - \left( \frac{x-\delta}{1-\delta} \right) \right)^{\frac{1}{P}}, \tag{3}$$

given

(1)  $y(x) = 0$ when $x - \delta < 0$,
(2)  $y(x) = 1 - \left( 1 - \left( \frac{x-\delta}{1-\delta} \right) \right)^{\frac{1}{P}}$ when $0 \le x - \delta \le 1$.

Note that, if we separately take the integral of the linear function, the exponential function, and the functional form implied by Pareto distribution from 0 to 1, it can be shown that each functional form has the same area under the Lorenz curve which equals $\frac{(1-\delta)}{(P+1)}$. We categorize these three functional forms into two components. The first component is the linear function and the second component is the weighted linear convex combination of the exponential function and the functional form implied by Pareto distribution, where the weight $(1 - \omega)$ is assigned to the exponential function and the weight $\omega$ is assigned to the functional form implied by Pareto distribution. As discussed in Introduction, the weighted linear combination of the exponential function and the functional form implied by Pareto distribution represents the convex segment of the Lorenz curve whereas the linear function characterizes the horizontal-line segment of the Lorenz curve where a certain number of observations have a size of zero and/or to represent the positive-slope linear segments of the Lorenz curve where the size of one observation is larger than the size of the others which are smaller and have approximately equal size. By assigning the weight $(1 - \rho)$ to the linear function and the weight $\rho$ to the weighted average of the exponential function and the functional form implied by Pareto distribution, the proposed model for the Lorenz curve can be shown as Eq. (4).

$$y(x) = (1-\rho) * \left[ \left( \frac{2}{P+1} \right) * \left( \frac{x-\delta}{1-\delta} \right) \right] + \rho * \left[ (1-\omega) * \left( \frac{x-\delta}{1-\delta} \right)^P + \omega * \left( 1 - \left( 1 - \left( \frac{x-\delta}{1-\delta} \right) \right)^{\frac{1}{P}} \right) \right]. \tag{4}$$

Our proposed model satisfies all necessary and sufficient conditions for the theoretical Lorenz curve which are as follows:

(1)  $y(0) = 0$,
(2)  $y(1) = 1$,
(3)  $\frac{dy}{dx} = (1-\rho) * \left[ \left( \frac{2}{P+1} \right) * \left( \frac{1}{1-\delta} \right) \right] + \rho * \left[ \frac{(1-\omega)*P*\left( \frac{x-\delta}{1-\delta} \right)^P}{(x-\delta)} + \omega * \left( \frac{1}{P} \right) * \frac{\left( \frac{1-x}{1-\delta} \right)^{\frac{1}{P}}}{(1-x)} \right]$

$\ge 0$, given $0 \le x - \delta \le 1$,

(4)  $\frac{d^2y}{dx^2} = \rho * \left[ (1-\omega) * \frac{(P-1)*P*\left( \frac{x-\delta}{1-\delta} \right)^P}{(x-\delta)} + \omega * \frac{\left( \frac{P-1}{P} \right)*\left( \frac{1}{P} \right)*\left( \frac{1-x}{1-\delta} \right)^{\frac{1}{P}}}{(1-x)^2} \right]$

$\ge 0$, given $0 \le x - \delta \le 1$.

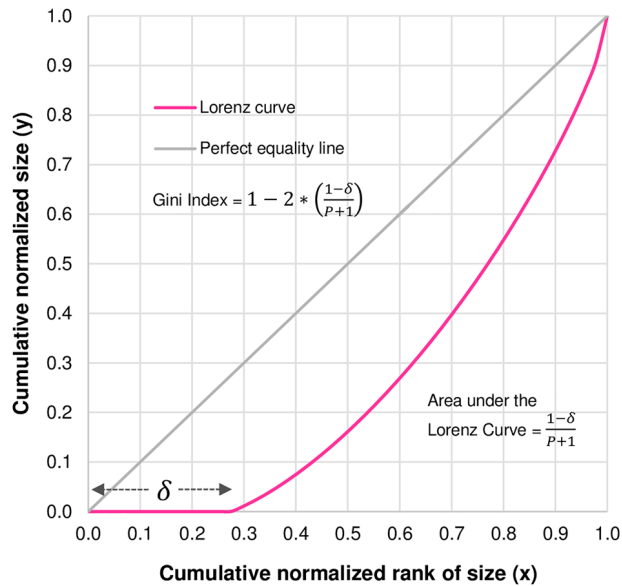

**Figure 1.** The Lorenz curve and the closed-form expression for the Gini index.

Note that if all necessary and sufficient conditions for the theoretical Lorenz curve as specified above are satisfied, the Lorenz curve is convex except when the parameters $\delta = 0$ and $P = 1$, the Lorenz curve is linear. Figure 1 illustrates the Lorenz curve that is consistent with the proposed model.

According to Sitthiyot and Holasut[34], different scientific disciplines may have their own theoretical justifications when applying the parametric functional form for the Lorenz curve to examine size distributions of nonnegative quantities and calculate statistical evenness measures. However, irrespective of disciplines of sciences, the parameter $\delta$ measures the distance along the horizontal-line segment of the Lorenz curve where the value of the cumulative normalized size $(y)$ is equal to zero for a given range of the cumulative normalized rank of size $(x)$ as illustrated in Fig. 1. As described above, the parameter $\rho$ is the weight given to the convex segment of the Lorenz curve while the weight $(1 - \rho)$ is given to the linear segment of the Lorenz curve. The parameter $P$ represents the degree of inequality or unevenness in size distribution as measured by the Gini index. The parameter $\omega$ is the weight that controls the curvature of the Lorenz curve such that the Gini index remains unchanged since, for a particular value of parameter $P$, there are infinite values of parameter $\omega$ that could give an identical value of the Gini index. The parameter $\omega$ thus provides the information about size shares in case two or more Lorenz curves intersect. In addition, from an analytical point of view, the key advantage of using the weighted linear convex combination of the exponential function and the functional form implied by Pareto distribution is that the shape of the estimated Lorenz curve could be handily adjusted via the change in parameter $\omega$ while the value of the Gini index is held constant. This may not be easily done for linear convex combinations of other functional forms for the Lorenz curve. To our knowledge, no study has employed a parametric functional form for estimating the Lorenz curve by combining the linear function, the exponential function, and the functional form implied by Pareto distribution before. The closest one is the model proposed by Sarabia[23] whose parametric functional form represents a linear convex combination of the egalitarian line, the power Lorenz curve, and the classical Pareto Lorenz curve. Based on our proposed model as shown in Eq. (4), the area under the Lorenz curve and the closed-form expression for the Gini index can be conveniently calculated as shown as Eqs. (5) and (6), respectively.

$$\int_0^1 y(x)dx = \frac{(1-\delta)}{(P+1)}. \tag{5}$$

$$\text{Gini index}_{\text{Proposed}} = 1 - 2 * \int_0^1 y(x)dx = 1 - 2 * \left(\frac{1-\delta}{P+1}\right), \tag{6}$$

$$0 \leq \text{Gini index}_{\text{Proposed}} \leq 1.$$

The Gini index takes values between 0 and 1. The closer the index is to 0, the more equal the distribution of size whereas the closer the index is to 1, the more unequal the size distribution. The formulae for calculating the area under the Lorenz curve and for computing the closed-form expression for the Gini index are also shown in Fig. 1.

To demonstrate the performance of the proposed model for the Lorenz curve, this study utilizes the data on sizes of events or things that occur in both nature and society from different scientific disciplines. In addition, we create a hypothetical dataset, representing a society that exhibits extreme inequality in income distribution in that 99 persons have an equal income of one unit and only one person has income of 99 units, in order to illustrate that our proposed model could be used to fit the distribution of size where one observation has a larger size than the others which have a smaller and equal size. The list of data and their sources are provided in Table 1.

| Data | Source |
|------|--------|
| The intensities of earthquakes occurring in California between 1910 and 1992 (Richter magnitude) | Newman[1] |
| Peak gamma-ray intensity of solar flares between 1980 and 1989 (scintillation counts per second) | Newman[1] |
| Body mass of late Quaternary mammals (grams) | Smith et al.[2] |
| The degrees of metabolites in the metabolic network of the bacterium *Escherichia coli* (number of connections) | Huss and Holme[3] |
| Compensation of CEOs of companies listed in the Standard & Poor's 500 in 2020 (United States dollar) | AFL-CIO[4] |
| Salary of professional women tennis players in 2019 (United states dollar) | Sitthiyot[5] |
| The intensities of inter-state wars (number of deaths per battle) | Sarkees and Wayman[6] |
| Hypothetical data (one person has income of 99 units while the other 99 persons have an equal income of one unit) | Authors' own data |

**Table 1.** The list of the data on sizes of events or things occurring in nature and society and the hypothetical data as well as their sources.

According to the proposed model as shown in Eq. (4), the parameters $\delta$, $\rho$, $\omega$, and $P$ can be estimated by using the curve fitting technique based on minimizing sum of squared errors. Let $e_i^2$ be the squared error, $y_i$ be the actual cumulative normalized size from the smallest to the largest, $\widehat{y_i}$ be the estimated cumulative normalized size from the smallest to the largest, and N be the number of observations. The minimization of sum of squared errors $\left(min \sum_{i=1}^{N} e_i^2\right)$ can be calculated as $min \sum_{i=1}^{N} \left(y_i - \widehat{y_i}\right)^2$. That is, for any given $x_i$, where $x_i$ is the cumulative normalized rank of size from the smallest to the largest, we have to find the values of parameters $\delta$, $\rho$, $\omega$, and $P$ such that $\sum_{i=1}^{N} \left(y_i - \widehat{y_i}\right)^2$ is minimized. Note that, given the estimated values of parameters $\delta$, $\rho$, $\omega$, and $P$, $\widehat{y_i}$ is computed from Eq. (4) by plugging in $x_i$.

To evaluate how well the estimated Lorenz curves fit both the empirical and the hypothetical data, this study employs five goodness-of-fit statistics which are coefficient of determination ($R^2$), mean squared error (MSE), mean absolute error (MAE), maximum absolute error (MAS), and information inaccuracy measure (IIM) developed by Theil[38] which can be computed as $\sum_{i=1}^{N} y_i * log_{10} \left(\frac{y_i}{\widehat{y_i}}\right)$. The closer the value of $R^2$ is to 1 as well as the closer the values of MSE, MAE, and MAS are to 0, the better the estimated functional form. For the IIM criterion, the estimated functional form that has a smaller absolute value of IIM is better than those with larger absolute values of IIM. In addition, we compare the performance of our proposed model to that of the SCS model which, according to Tanak et al.[37], has the best overall performance among a number of different functional forms employed in approximating the Lorenz curve. By using the same notations for the cumulative normalized rank of size ($x$) and the cumulative normalized size ($y$) as before and also letting $\gamma$, $\alpha$, and $\beta$ denote parameters, where $\gamma \geq 0$, $0 < \alpha \leq 1$, and $\beta \geq 1$, the SCS model is shown as Eq. (7).

$$y(x) = x^\gamma * \left(1 - (1-x)^\alpha\right)^\beta. \tag{7}$$

Note that the SCS model does not have an explicit mathematical solution for the Gini index. Therefore, we calculate the value of the estimated Gini index based on the SCS model (Gini index$_{SCS}$) by using the numerical integration. In addition to the SCS model, we compare the performance of our proposed model to that of the S model since it has four parameters and is shown to outperform other well-known parametric functional forms for estimating the Lorenz curve as discussed in Introduction. For notations, let $\lambda$, $\eta$, $a_1$ and $a_2$ denote parameters, where $a_1 \geq 0$, $a_2 + 1 > 0$, $\eta * a_2 + \lambda \leq 1$, $\lambda \geq 0$, and $\eta * a_2 \geq 0$, the S model can be shown as Eq. (8).

$$y(x) = (1 - \lambda + \eta) * x + \lambda * x^{a_1+1} - \eta * \left[1 - (1-x)^{a_2+1}\right]. \tag{8}$$

According to Sarabia[23], the S model has a closed-form expression for the Gini index which is shown as Eq. (9).

$$\text{Gini index}_S = \lambda * \left(1 - \frac{2}{2+a_1}\right) + \eta * \left(1 - \frac{2}{2+a_2}\right). \tag{9}$$

This study uses the Microsoft Excel Data Analysis program and the Microsoft Excel Solver program, which are available in most, if not all, computers, for calculating the descriptive statistics as well as estimating the parameters and calculating the values of estimated Gini index. As suggested by Dagum[35], from a viewpoint of computational cost and the acceptance of the specified functional form in applied sciences, a simple method of parameter estimation is always an advantage. Table 2 reports the descriptive statistics of the data on sizes of events or things occurring in nature and society as well as the hypothetical data representing the case where one observation is larger in size (one person has income of 99 units) compared to the rest of observations which are smaller and equal in size (99 persons have an equal income of one unit).

| Size | Minimum | Maximum | Mean | Standard deviation | Number of observations (N) |
|---|---|---|---|---|---|
| Earthquake intensity (Richter magnitude) | 0 | 7.80 | 2.90 | 1.18 | 19,302 |
| Solar flare intensity (scintillation counts per second) | 20.00 | 231,300.00 | 689.41 | 6,520.59 | 12,773 |
| Mammal body mass (grams) | 1.75 | 190,000,000.00 | 182,341.15 | 3,402,944.82 | 4910 |
| Metabolic degree (number of connections) | 0 | 485.00 | 3.51 | 30.75 | 468 |
| CEOs' compensation (United States dollar) | 1.00 | 211,131,206.00 | 15,532,683.63 | 15,005,270.77 | 498 |
| Salary of professional women tennis players (United States dollar) | 37.00 | 2,916,508.00 | 27,520.67 | 122,838.31 | 968 |
| Inter-state war intensity (number of deaths per battle) | 0 | 7,500,000.00 | 95,478.90 | 501,920.17 | 336 |
| Hypothetical data (one person has income of 99 units while the other 99 persons have an equal income of one unit) | 1.00 | 99.00 | 1.98 | 9.80 | 100 |

**Table 2.** The descriptive statistics of the empirical data on sizes of events or things occurring in nature and society and the hypothetical data.

| Size | Estimated parameters | | | | | | | | | | |
|---|---|---|---|---|---|---|---|---|---|---|---|
| | Proposed | | | | SCS | | | S | | | |
| | $\delta$ | $\rho$ | $\omega$ | $P$ | $\gamma$ | $\alpha$ | $\beta$ | $\lambda$ | $\eta$ | $a_1$ | $a_2$ |
| Earthquake intensity | 0.10 | 1.00 | 0.49 | 1.29 | 0.00 | 0.99 | 1.49 | 0.98 | 8.72 | 0.49 | 0.00 |
| Solar flare intensity | 0.00 | 1.00 | 0.91 | 15.34 | 0.00 | 0.14 | 1.39 | 0.77 | 42.45 | 186.62 | 0.00 |
| Mammal body mass | 0.00 | 1.00 | 0.53 | 96.93 | 0.00 | 0.37 | 8.52 | 0.97 | 10.70 | 291.28 | 0.00 |
| Metabolic degree | 0.90 | 0.88 | 0.00 | 13.42 | 0.00 | 0.91 | 95.82 | 0.99 | 5.20 | 153.16 | 0.00 |
| CEOs' compensation | 0.00 | 1.00 | 0.61 | 2.01 | 0.00 | 0.67 | 1.29 | 0.04 | 68.42 | 341.63 | 0.01 |
| Salary of professional women tennis players | 0.00 | 1.00 | 0.29 | 13.56 | 0.00 | 0.62 | 5.61 | 0.89 | 242.04 | 18.73 | 0.00 |
| Inter-state war intensity | 0.58 | 1.00 | 0.39 | 11.62 | 0.00 | 1.00 | 34.83 | 0.92 | 202.97 | 42.48 | 0.00 |
| Hypothetical data | 0.00 | 0.00 | 0.00 | 2.96 | 0.00 | 0.31 | 1.00 | 0.00 | 34,887.31 | 2.16 | 0.00 |

**Table 3.** The estimated parameters based on the proposed model, the SCS model, and the S model.

| Size | Goodness-of-fit statistics | | | | | | | | | |
|---|---|---|---|---|---|---|---|---|---|---|
| | $R^2$ | | MSE | | MAE | | MAS | | IIM | |
| | Proposed | SCS | Proposed | SCS | Proposed | SCS | Proposed | SCS | Proposed | SCS |
| Earthquake intensity | **1.0000** | 0.9987 | **0.0165** | 2.3022 | **0.0008** | 0.0086 | **0.0022** | 0.0299 | **0.9095** | 12.3513 |
| Solar flare intensity | 0.9296 | 0.9509 | 5.2676 | 3.6752 | **0.0070** | 0.0100 | 0.4635 | 0.3376 | 40.2341 | 30.6602 |
| Mammal body mass | 0.9527 | 0.9976 | 0.3839 | 0.0196 | 0.0047 | 0.0004 | 0.2782 | 0.0930 | 1.0813 | 0.6811 |
| Metabolic degree | 0.9924 | 0.9963 | **0.0057** | 0.0075 | **0.0009** | 0.0011 | 0.0285 | 0.0279 | **0.1569** | 0.4789 |
| CEOs' compensation | **0.9995** | 0.9994 | **0.0170** | 0.0203 | 0.0049 | 0.0049 | **0.0145** | 0.0240 | 0.3590 | 0.3439 |
| Salary of professional women tennis players | 0.9952 | 0.9964 | 0.0965 | 0.0719 | 0.0079 | 0.0050 | 0.0732 | 0.0582 | 1.0503 | 0.0921 |
| Inter-state war intensity | **0.9969** | 0.9438 | **0.0125** | 0.2247 | **0.0028** | 0.0131 | **0.0697** | 0.1551 | **0.1481** | 6.5223 |
| Hypothetical data | **1.0000** | 0.8586 | **0.0000** | 0.3698 | **0.0000** | 0.0474 | **0.0000** | 0.2589 | **0.0000** | 1.0848 |

**Table 4.** The evaluation of performance of the proposed model and that of the SCS model based on five goodness-of-fit statistics. The bold numbers indicate that the proposed model is superior to the SCS model.

## Results and discussion

Table 3 reports the results of the estimated parameters based on the proposed model, the SCS model, and the S model. We first compare the performance of our proposed model to that of the SCS model. As shown in Table 4, the values of $R^2$ ranging between 0.9296 and 1.000 for the proposed model and between 0.8586 and 0.9994 for the SCS model suggest that all estimated Lorenz curves fit the empirical and the hypothetical data reasonably well. While both models perform equally well on the criteria of $R^2$, MAS, and IIM, our proposed model slightly outperforms the SCS model on the basis of MSE and MAE.

Considering the earthquake intensity and the inter-state war intensity whose sizes contain zeros, the results, as reported in Table 4, indicate that our proposed model outperforms the SCS model in all five goodness-of-fit statistics while it outperforms the SCS model in three out of five statistical measures of goodness-of-fit for the metabolic degree. Moreover, our proposed model fits the hypothetical data perfectly well whereas the SCS model fits the hypothetical data relatively less well on the criteria of $R^2$, MSE, MAE, MAS, and IIM.

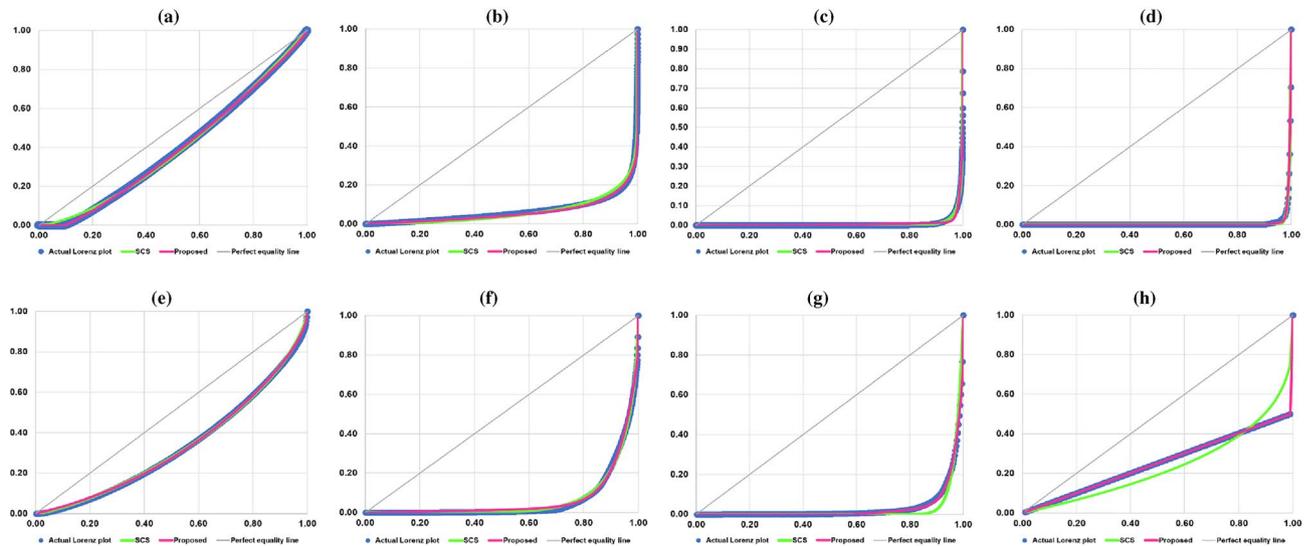

**Figure 2.** The actual Lorenz plots and the estimated Lorenz curves based on the proposed model and the SCS model. (**a**) Earthquake intensity. (**b**) Solar flare intensity. (**c**) Mammal body mass. (**d**) Metabolic degree. (**e**) CEOs' compensation. (**f**) Salary of professional women tennis players. (**g**) Inter-state war intensity. (**h**) Hypothetical data.

| | Goodness-of-fit statistics | | | | | | | | | |
|---|---|---|---|---|---|---|---|---|---|---|
| | $R^2$ | | MSE | | MAE | | MAS | | IIM | |
| Size | Proposed | S | Proposed | S | Proposed | S | Proposed | S | Proposed | S |
| Earthquake intensity | **1.0000** | 0.9987 | **0.0165** | 2.2905 | **0.0008** | 0.0085 | **0.0022** | 0.0299 | **0.9095** | 12.2121 |
| Solar flare intensity | 0.9296 | 0.9699 | 5.2676 | 2.2507 | **0.0070** | 0.0084 | 0.4635 | 0.1253 | 40.2341 | 12.2970 |
| Mammal body mass | **0.9527** | 0.8119 | **0.3839** | 1.5264 | **0.0047** | 0.0081 | 0.2782 | 0.2436 | **1.0813** | 4.3338 |
| Metabolic degree | 0.9972 | 0.9966 | **0.0057** | 0.0069 | **0.0009** | 0.0021 | 0.0285 | 0.0202 | **0.1569** | 0.1719 |
| CEOs' compensation | **0.9995** | 0.9987 | **0.0170** | 0.0450 | **0.0049** | 0.0085 | **0.0145** | 0.0150 | **0.3590** | 0.4722 |
| Salary of professional women tennis players | **0.9952** | 0.9646 | **0.0965** | 0.7142 | **0.0079** | 0.0151 | **0.0732** | 0.1550 | **1.0503** | 1.1846 |
| Inter-state war intensity | 0.9969 | 0.9740 | **0.0125** | 0.1041 | **0.0028** | 0.0103 | **0.0697** | 0.1289 | **0.1481** | 0.2182 |
| Hypothetical data | **1.0000** | 0.3618 | **0.0000** | 1.6693 | **0.0000** | 0.0923 | **0.0000** | 0.4473 | **0.0000** | 0.5818 |

**Table 5.** The evaluation of performance of the proposed model and that of the S model based on five goodness-of-fit statistics. The bold numbers indicate that the proposed model is superior to the S model.

In total, there are 22 out of 40 cases where the proposed model outperforms the SCS model while there are 17 cases in which the SCS model performs better than the proposed model. There is one case which is the size of CEOs' compensation where both models perform equally well on the basis of MAE. Thus, on the criteria of five statistical measures of goodness-of-fit, namely, $R^2$, MSE, MAE, MAS, and IIM, we can conclude that the performance of our proposed model is slightly better than that of the SCS model. Nonetheless, the proposed model is clearly superior to the SCS model when the data contain zeros and/or exhibit extreme inequality. Figure 2 shows the actual Lorenz plots of the data on sizes and their corresponding estimated Lorenz curves based on the proposed model and the SCS model.

Next, we report the performance comparison between the proposed model and the S model. The results are shown in Table 5. On the basis of $R^2$, MSE, MAE, MAS, and IIM, our proposed model outperforms the S model in 33 out of 40 cases. Focusing on the cases where the data contain zeros, the proposed model outperforms the S model for the earthquake intensity and the inter-state war intensity as measured by all five goodness-of-fit statistics while it outperforms the S model in three out of five goodness-of-fit statistics for the metabolic degree. In addition, the values of $R^2$, MSE, MAE, MAS, and IIM indicate that our proposed model fit the hypothetical data better than the S model. Figure 3 illustrates the actual Lorenz plots of the data on sizes and their corresponding estimated Lorenz curves according to the proposed model and the S model.

The overall performance comparison between the proposed model and the SCS model and that between the proposed model and the S model indicate that, on the basis of $R^2$, MSE, MAE, MAS, and IIM, our proposed model, by and large, is superior to the SCS model and the S model, especially when the data contain zeros and/or exhibit extreme inequality. For the data containing zeros, this can be demonstrated by the positive values of parameter $\delta$ as shown in Table 3 and also the horizontal-line segments of the estimated Lorenz curves for the earthquake intensity, the metabolic degree, and the inter-state war intensity as illustrated in Figs. 2a,d,g and 3a,d,g. In addition, for the data exhibiting extreme inequality where one observation has a larger size (one person has income of 99 units) than the others which have smaller and equal size (99 persons have an equal income of one unit), the proposed model would be able to perfectly fit the hypothetical data which can be demonstrated by the two positive-slope linear segments as shown in Figs. 2h and 3h while the SCS model and the S model fall short of this task.

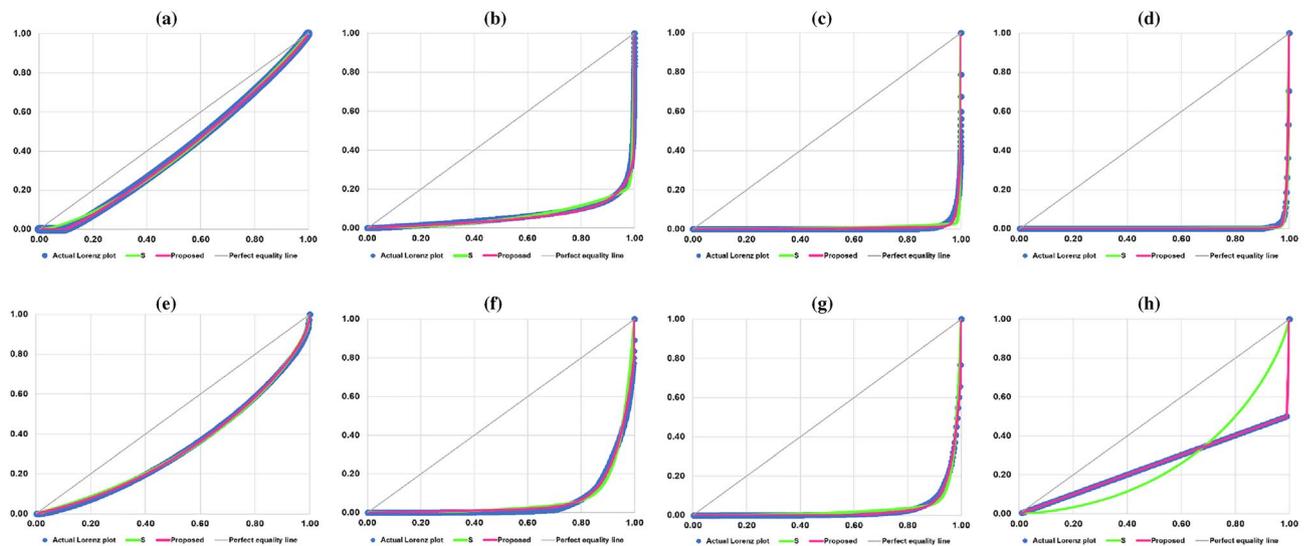

**Figure 3.** The actual Lorenz plots and the estimated Lorenz curves based on the proposed model and the S model. (**a**) Earthquake intensity. (**b**) Solar flare intensity. (**c**) Mammal body mass. (**d**) Metabolic degree. (**e**) CEOs' compensation. (**f**) Salary of professional women tennis players. (**g**) Inter-state war intensity. (**h**) Hypothetical data.

Furthermore, our proposed model and the S model have a closed-form expression for the Gini index which can be conveniently computed by using Eqs. (6) and (9) as shown in "Methods". The SCS model, however, requires the valuations of the beta and the gamma functions or the numerical integration in order to estimate the Gini index since its explicit mathematical solution for the Gini index does not exist. Table 6 reports the values of the estimated Gini index based on the proposed model, the SCS model, and the S model. Note that, for the SCS model, we calculate the estimated Gini index by using the numerical integration.

The results, as reported in Table 6, suggest that the values of the estimated Gini index calculated based on the proposed model do not significantly differ from those computed based on the SCS model and the S model except for the case of extreme inequality where the proposed model perfectly fits the hypothetical data which results in the value of estimated Gini index being identical to its actual value which is equal to 0.495. The estimated Gini index calculated based on the SCS model and the S model, however, are equal to 0.527 and 0.464 which differ from the actual Gini index by 0.032 and 0.031, respectively.

Even though one of the objectives for developing a model for the Lorenz curve is to calculate the Gini index that would be close to the actual observation, this study would like to note that, when assessing the performance of different parametric functional forms, the goodness-of-fit statistical measures, the shape of the estimated Lorenz curve, and the estimated Gini index should be taken into consideration altogether since there are infinite number of the Lorenz curves that could result in the same value of the Gini index. Thus, a good model for the Lorenz curve must be able to describe the shape of the distribution of size through changes in the values of parameters and the fact that it fits the actual data would be the main reason for its choice[35].

## Conclusions

Given that previous studies have shown that no parametric functional form for the Lorenz curve is always optimal, different attempts therefore are still worth studying[39]. This study introduces a universal model for the Lorenz curve with an explicit mathematical solution for the Gini index. By using the empirical datasets on sizes of events or things occurring in both nature and society from different disciplines of sciences, some of which contain zeros, as well as the hypothetical dataset created in order to represent the situation where one observation has a larger size than the rest of observations which have smaller and equal size, this study demonstrates that the proposed model fits not only the data whose actual Lorenz plots are convex but also the data whose actual Lorenz plots are both horizontal and convex practically well. It also fits the distribution of size where one observation has a larger size compared to the others which have smaller and equal size as characterized by the estimated Lorenz curve that has two positive-slope linear segments. To our knowledge, no study has proposed a parametric functional form for the Lorenz curve that could fit the data that have a typical convex segment, a horizontal-line segment and a convex segment, or two positive-slope linear segments before.

To evaluate the performance of our universal model for the Lorenz curve, this study compares the performance of the proposed model to those of the SCS model and the S model, both of which have been shown to outperform other well-known parametric functional forms for the Lorenz curve[23,37]. The results indicate that the proposed model, by and large, is superior to both the SCS model and the S model on the criteria of $R^2$, MSE, MAE, MAS, and IIE, especially when the datasets contain zeros and/or exhibit extreme inequality in that one observation has a larger size than the rest of observations which have smaller and equal size. The other advantage is that the estimated Gini index based on our proposed model is more convenient to calculate than that computed based on the SCS model. This is because our proposed model has a closed-form expression for the Gini index whereas an explicit mathematical solution for the Gini index based on the SCS model does not exist and requires the valuations of the beta and gamma functions or the numerical integration. Moreover, when the dataset contains one observation which has a larger size than the rest of observations which have smaller and equal size, the estimated Gini index calculated based on our proposed model is identical to the actual Gini index whereas those calculated based on the SCS model and the S model are a bit of the mark.

| Size | Estimated Gini index | | | Absolute value of the difference between proposed and SCS | Absolute value of the difference between proposed and S |
|---|---|---|---|---|---|
| | Proposed | SCS | S | | |
| Earthquake intensity | 0.212 | 0.204 | 0.204 | 0.007 | 0.007 |
| Solar flare intensity | 0.878 | 0.872 | 0.867 | 0.005 | 0.011 |
| Mammal body mass | 0.980 | 0.984 | 0.977 | 0.005 | 0.002 |
| Metabolic degree | 0.986 | 0.986 | 0.983 | 0.000 | 0.003 |
| CEOs' compensation | 0.336 | 0.337 | 0.335 | 0.000 | 0.002 |
| Salary of professional women tennis players | 0.863 | 0.871 | 0.860 | 0.008 | 0.003 |
| Inter-state war intensity | 0.933 | 0.944 | 0.920 | 0.011 | 0.013 |
| Hypothetical data | 0.495 | 0.527 | 0.464 | 0.032 | 0.031 |

**Table 6.** The comparison of values of the estimated Gini index calculated based on the proposed model, the SCS model, and the S model.

Considering that the Lorenz curve and the Gini index are widely used in many scientific disciplines, we hope that our universal model for the Lorenz curve with a closed-form expression for the Gini index could be useful for analyzing the distributions of sizes and investigating their inequalities or unevennesses.

## Data availability

All data analyzed during this study are publicly available and can be accessed from the sources listed in Table 1 and also in References. Note that while the original sources of data on the earthquake intensity, the solar flare intensity, and the metabolic degree are provided in Table 1 and in References, this study obtained all three datasets from Clauset et al.[40] who also use them in their study. These three datasets are available at https://aaronclauset.github.io/powerlaws/data.htm.

## Acknowledgements

The authors are grateful to Dr. Suradit Holasut for guidance and comments.

## Author contributions

T.S. conceived the study. T.S. designed the methodology and performed the analysis. K.H. validated the results. T.S. wrote the main manuscript text. T.S. and K.H. reviewed and edited the main manuscript text. Both authors reviewed the manuscript.

## Funding

This research did not receive any specific grant from funding agencies in the public, commercial, or not-for-profit sectors.

## Competing interests

The authors declare no competing interests.

## Additional information

**Correspondence** and requests for materials should be addressed to T.S.

**Reprints and permissions information** is available at www.nature.com/reprints.

**Publisher's note**  Springer Nature remains neutral with regard to jurisdictional claims in published maps and institutional affiliations.